# Fast Ion Surface Energy Loss and Straggling in the Surface Wake Fields

T. Nandi,[1,*] K. Haris,[2] Hala,[2] Gurjeet Singh,[3] Pankaj Kumar,[1] Rajesh Kumar,[1] S. K. Saini,[1]
S. A. Khan,[1] Akhil Jhingan,[1] P. Verma,[4] A. Tauheed,[2] D. Mehta,[3] and H. G. Berry[5,#]

[1]Inter-University Accelerator Centre, JNU New Campus, New Delhi 110067, India
[2]Department of Physics, Aligarh Muslim University, Aligarh 202002, India
[3]Department of Physics, Panjab University, Chandigarh 160014, India
[4]Department of Physics, Kalindi College, East Patel Nagar, New Delhi 110008, India
[5]Department of Physics, University of Notre Dame, Notre Dame, Indiana 46556, USA

*nanditapan@gmail.com  #hgberry@nd.edu

We have measured the stopping powers and straggling of fast, highly ionized atoms passing through thin bilayer targets made up of metals and insulators. We were surprised to find that the energy losses as well as the straggling depend on the ordering of the target and have small but significantly different values on bilayer reversal. We ascribe this newly found difference in energy loss to the surface energy loss field effect due to the differing surface wake fields as the beam exits the target in the two cases. This finding is validated with experiments using several different projectiles, velocities, and bilayer targets. Both partners of the diatomic molecular ions also display similar results. A comparison of the energy loss results with those of previous theoretical predictions for the surface wake potential for fast ions in solids supports the existence of a self-wake.

PACS numbers: 34.35.+a, 34.20. b

Energy dissipation of fast charged particles through matter has been a subject of great interest for 100 years [1]. Although the energy loss mechanism in solids consists of contributions from both the bulk as well as the surfaces of a thin target foil [2], most studies consider only the bulk effects because major energy loss takes place through ionization processes in the bulk. Much smaller dissipating channels such as excitation and charge exchange processes can occur in both the bulk and at the surface. The energy loss contribution from the bulk is presumed to be much larger than that from the surface, although sometimes even the energy loss at the front surface can supersede the bulk energy loss for highly charged slow ions in very thin solid foils [3].

Besides the excitation and the charge exchange processes at the surface, three other processes can be responsible for the energy loss: the ion interaction with the surface potential barrier [4], with the image potential [5], and with the wake potential [6]. The first process is important only at the front surface. The second exists at both surfaces but its magnitude is very small for ion velocities higher than the Fermi velocity ($v_f$) of the electrons in the target. The third acts only at the exit surface and is significant at high velocities. For slow grazing incidence ions [7,8] it is not possible to segregate the surface effects from the bulk. For experiments conducted with ions traversing a target, the bulk energy loss will be admixed with the energy losses at the two surfaces. Segregating the contributions of the two surfaces from the bulk has not yet been possible. However, many years ago, bulk wake-field-induced Stark mixing of the sub states in H-like Kr ions [9] and recently the surface wake field intensity [10] in carbon foils have been measured. The latter showed that small surface wake field can be distinguished from the large bulk energy loss field in an atomic level lifetime measurement [10]. This experiment in fact supports the original Bohr prediction [11]. In this Letter, we search for the effect of the exit-surface wake potential through direct energy loss measurements. Our challenge has been to develop an energy loss measurement technique which can distinguish the bulk and surface energy loss contributions: we have achieved this by the simple trick of using bilayer reversible targets.

All theoretical developments [12–15] assume that the wake potential is caused by collective plasmon excitations of the electronic states of the fast beam ions formed at the exit surface of a conducting foil. Our hypothesis was that in the case of a bilayer target (one part metallic and other part insulating), the bulk wake will be formed equally in both configurations, but, the surface wake will be greater for the beam inputting on the insulator side and exiting from the metal side of the bilayer target. Thus, the differences in the wake fields for two different orientations could be large enough to become measurable: the inverse geometry with the insulator at the exit side of the target foil will not allow much collective plasmon excitation to take place. As a result, in one configuration the wake potential is present and in the other it is absent, and thus the energy loss difference between the two geometries should give a measure of both the ion energy loss and the straggling in the surface wake potential: we have identified this as the surface energy loss field (SELF) effect. The SELF is not merely important in understanding ion-matter interaction but also it finds applications in broad areas of radiation damage, including biological systems. Aluminized

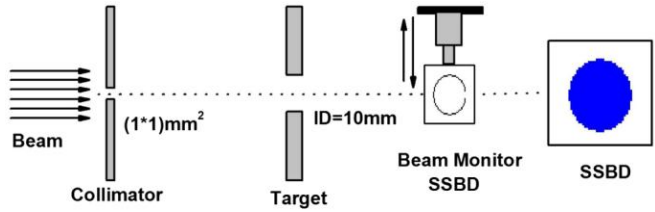

FIG. 1 (color online). Schematic diagram of the experimental arrangement.

polypropylene is an important window foil for gas detectors. Metal semiconductor bilayers are important in applications of biological detection, nanothermometers, photo catalysis, photocells, and detectors [16].

The present measurements have been performed using the 15 UD (unit doubled) tandem Pelletron accelerator at the Inter-University Accelerator Center [17]. Ion beams of $^{56}$Fe with energy ranging 85–155 MeV and molecular ion beam of $^{16}$OH$^+$ of energy 20 MeV were passed through various bilayer targets, viz., polypropylene and aluminum (PP-Al), polyethylene terephthalate and germanium (PET-Ge), and polypropylene and gold (PP-Au). The schematic of the experimental setup is shown in Fig. 1. Quartz glass and bilayer targets were mounted on the target manipulator system, which was able to move in the y direction (perpendicular to the z-axis beam direction) and about the y axis too. Rotating and then anchoring it with two bolts ensures each bilayer target flipped by exactly 180º rotation in the experiment. Three empty positions were retained in the ladder so that the beam energy measurements could be repeated six times. The ion beam is incident directly on the detector through the bilayer target. In order to minimize the electrical noise, an electrically isolated system with a good grounding connection was used for the detector and the electronics. Further, the detector was cooled up to -25º C using a Peltier element to minimize the thermal noise. This helped to achieve good resolution of the system (20 keV at 5.48 MeV alpha) and prevent damage to the detector.

The beam intensity was required to be significantly reduced to avoid damage to the detector. At the first step, a quality ion beam was prepared and collimated through a double slit device allowing only a reduced beam of size 1 mm x 1 mm. This facilitates in keeping the beam optical parameters constant. In order to view and monitor the beam intensity, the beam was first tuned on a piece of quartz placed on the target ladder. A satisfactory beam shape and beam transmission condition of the accelerator from the sputtered negative ion cesium source (SNICS) was obtained at the target position. The SNICS parameters such as oven temperature and cesium focus were reduced to a certain level so that the beam spot on quartz had totally disappeared. In the next step, a blank target position was brought to the beam path and direct beam was put for a short period onto a monitor detector by means of a pneumatically controlled linear motion vacuum feed through.

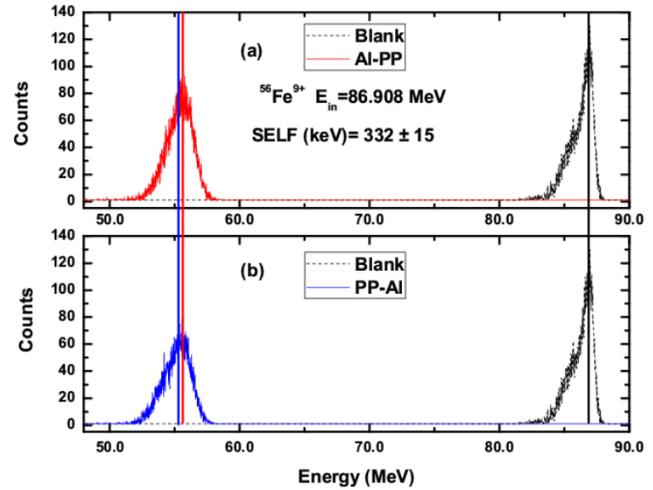

FIG. 2 (color online). Geometry dependent (forward and back) spectra, showing the differential surface energy loss field (SELF) effect of the bilayer targets (a) Al-PP and (b) PP-Al.

The ion source parameters were tuned to achieve the count rate in the monitor detector to ~50 counts/ sec . The monitor detector was moved out and the beam energy was measured using a good quality Peltier-cooled Si-surface barrier detector (SSBD).

Particle spectra were taken with and without the target foils and the energy differences obtained to give the energy losses of the particular ions through the target. The target surfaces were fixed at 90º to the beam direction throughout the experiment. The spectrum for the Al-PP bi- layer target (ion beam entering the Al side) is shown in Fig. 2(a). The corresponding spectrum for the PP-Al bilayer target is shown in Fig. 2(b). Using the solvent and weighing method, the thicknesses of PP and Al were found to be $6090.00 \pm 5.50$ nm and $30.35 \pm 0.74$ nm, respectively.

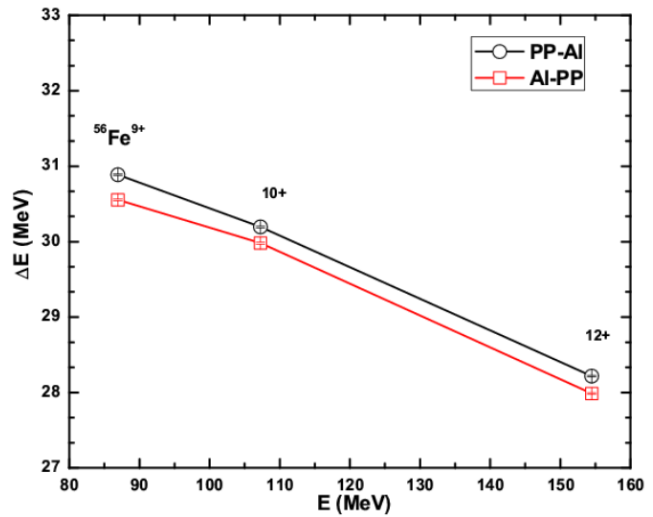

FIG. 3 (color online). Beam energy dependent energy loss in Al-PP and PP-Al bilayer targets illustrating the SELF difference effect.

TABLE I. Measured surface energy loss field (SELF) differences for different bilayer targets and estimates of the range of the surface wake field (SWF) at Al, Ge, and Au surfaces. $-dE/dx$ is estimated using the formula from Ref. [18].

| Ion species | Incident beam energy (MeV) | Energy loss ($\Delta E$) (MeV) | | Difference in $\Delta E$ (keV) | ($-dE/dx$) (keV/Å) | Range of SWF (Å) |
|---|---|---|---|---|---|---|
| | | PP-Al | Al-PP | | | Al-side |
| $^{56}Fe^{9+}$ | 86.908 ± 0.005 | 30.887 ± 0.011 | 30.555 ± 0.010 | 332 ± 15 | 7.41 | 45 ± 2 |
| $^{56}Fe^{10+}$ | 107.258 ± 0.006 | 30.193 ± 0.011 | 29.981 ± 0.012 | 212 ± 16 | 5.85 | 36 ± 3 |
| $^{56}Fe^{12+}$ | 154.473 ± 0.006 | 28.219 ± 0.009 | 27.988 ± 0.009 | 231 ± 13 | 3.98 | 58 ± 3 |
| | | PET-Ge | Ge-PET | | | Ge side |
| $^{56}Fe^{9+}$ | 86.908 ± 0.005 | 39.951 ± 0.014 | 39.541 ± 0.013 | 410 ± 19 | 5.64 | 73 ± 3 |
| $^{56}Fe^{10+}$ | 107.258 ± 0.006 | 39.120 ± 0.012 | 38.923 ± 0.013 | 197 ± 18 | 4.30 | 46 ± 4 |
| $^{56}Fe^{12+}$ | 154.473 ± 0.006 | 37.018 ± 0.014 | 36.841 ± 0.011 | 177 ± 17 | 2.80 | 63 ± 6 |
| | | PP-Au | Au-PP | | | Au side |
| $^{56}Fe^{9+}$ | 86.908 ± 0.005 | 29.340 ± 0.012 | 29.222 ± 0.012 | 118 ± 17 | 2.38 | 50 ± 7 |
| | | PP-Al | Al-PP | | | Al side |
| $^{16}O^{q+}$ | 18.824 ± 0.003 | 7.426 ± 0.001 | 7.400 ± 0.002 | 26 ± 2 | 0.89 | 29 ± 2 |
| $^{1}H^{+}$ | 1.176 ± 0.0002 | 0.409 ± 0.0002 | 0.407 ± 0.0003 | 2 ± 0.4 | 0.01 | 200 ± 36 |

Similar measurements for the energy loss have been performed for other bilayer targets of PET (6μm)-Ge (185 nm) and PP (6μm)-Au (15 nm).

The outgoing energies through the target foils were determined at all the beam energies with respect to the energy calibration through the blank position. The centroid of the outgoing energy peak through Al-PP, PP-Al and blank is marked in Fig. 2 for the representative case with 86.908 MeV $^{56}Fe^{9+}$ ions. The same experiment was carried out with 107.258 MeV $^{56}Fe^{10+}$, and 154.473 MeV $^{56}Fe^{12+}$. Each centroid (C) was determined from $C = \sum_{i=1}^{N} C_i n_i / \sum_{i=1}^{n} n_i$, where $C_i$ is the channel number and $n_i$ is corresponding count and its derivative gives a measure of its uncertainty.

The order-dependent energy losses are shown in Fig. 3 and the values are given in Table I. It can be observed immediately in the Table I that the energy losses for the PP-Al targets are more than the reversed Al-PP targets. The difference in energy loss is significant. Having observed the bilayer target order differential SELF energies for Al-PP targets, we made similar measurements with a few other targets such as PET-Ge and PP-Au and observed similar trends.

In order to understand the above differential energy loss differences, we performed theoretical calculations using the SRIM [19] and ATIMA codes [20] for a bilayer target of PP and Al for the two different geometries. The SRIM calculation shows a difference up to a few tens of keV depending on the incident energy, whereas ATIMA shows only a few keV energy losses. In both cases, the energy loss is higher in Al-PP than that in PP-Al as shown in Table II. The scenario was alike for Ge-PET and Au-PP too. Predictions from both the theories are consistent but their picture is opposite to those of the measurements. Hence the SELF differential energy loss cannot be explained by the present theories.

Passage of molecular ion beams through thin bilayer foils shows a difference with the molecular orientation axis relative to the beam direction, known as the

TABLE II. Theoretical comparisons for the SELF energy loss differences for Al-PP, Ge-PET, and Au-PP bilayer targets.

| Ion species | Incident beam energy (MeV) | ATIMA | | | SRIM | | |
|---|---|---|---|---|---|---|---|
| | | E (MeV) | | Difference in $\Delta E$(keV) | E (MeV) | | Difference in $\Delta E$(keV) |
| | | Al-PP | PP-Al | | Al-PP | PP-Al | |
| $^{56}Fe^{9+}$ | 86.908 ± 0.005 | 26.654 | 26.647 | 7 | 29.472 | 29.449 | 23 |
| $^{56}Fe^{10+}$ | 107.258 ± 0.006 | 26.182 | 26.178 | 4 | 28.144 | 28.124 | 20 |
| $^{56}Fe^{12+}$ | 154.473 ± 0.006 | 24.275 | 24.273 | 2 | 26.879 | 26.873 | 6 |
| | | Ge-PET | PET-Ge | | Ge-PET | PET-Ge | |
| $^{56}Fe^{9+}$ | 86.908 ± 0.005 | 37.395 | 37.196 | 199 | 37.958 | 37.787 | 171 |
| $^{56}Fe^{10+}$ | 107.258 ± 0.006 | 37.038 | 36.906 | 132 | 37.216 | 37.091 | 125 |
| $^{56}Fe^{12+}$ | 154.473 ± 0.006 | 34.765 | 34.706 | 59 | 35.075 | 35.041 | 34 |
| | | Au-PP | PP-Au | | Au-PP | PP-Au | |
| $^{56}Fe^{9+}$ | 86.908 ± 0.005 | 26.656 | 26.627 | 29 | 29.575 | 29.533 | 42 |

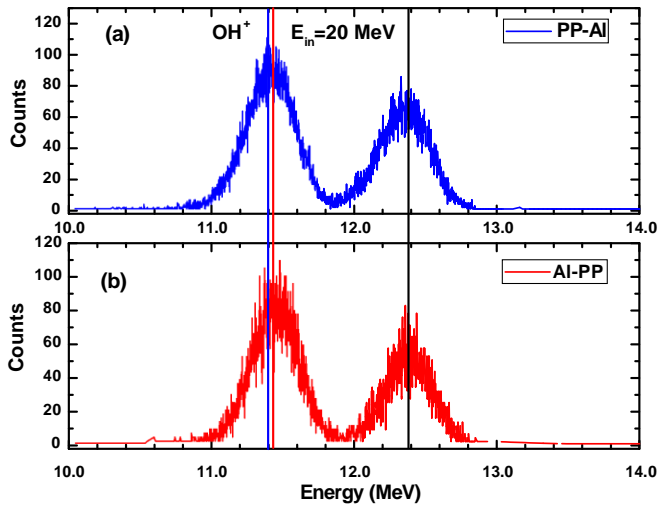

FIG. 4 (color online). Geometry dependent (forward and backward) energy loss spectra of the bilayer targets (a) PP-Al and (b) Al-PP for the molecular Coulomb explosion of $OH^+$ ion.

Coulomb explosion effect [21], and is attributed to the energy loss of the heavier ions in the polarization wakes of lighter ions. In the second part of these experiments, we used a 20 MeV $^{16}OH^+$ beam on Al-PP bilayer targets to observe the SELF differential energy losses - see Table I and Fig. 4. We measure SELF energy loss differences of $26 \pm 2$ and $2.0 \pm 0.2$ keV for oxygen and hydrogen ions, respectively. The former results thus reveal the effect of the surface polarization wake. But the latter, the energy loss difference of hydrogen ions, is a surprise. Convoy electrons produced due to passage of swift ions through solid target give rise to forming plasmon excitations in both the bulk of the target foil as well as at its surface. In particular, convoy electrons traveling with little higher speed than the ion might create a wake near the surface that every ion has to cross over. One such a wake is termed as the self-wake [15] and it could be responsible for the energy loss difference of hydrogen ions.

The average energy loss by an ion in a potential is the product of the effective charge state [6] times the potential itself. Thus, the measured energy loss from the surface wake potential divided by the effective charge state will provide its magnitude, including the potential trough and the self-wake. The magnitude of the calculated wake potential from Neelavathi et al. [13] is higher than that calculated from Refs. [6,14]. We have scaled the Neelavathi wake potential for O and S ions to our case of Fe ions. the wake potential turns out to be only about 160 V. In contrast, our measured wake potential is about 15 kV for PP-Al bilayers, which is at about 2 orders of magnitude higher.

Let us consider an alternative formula of Vager and Gemmell [18] for the stopping power. The bulk plasmon frequencies do not differ much on reversing the target order. However, the surface plasmon frequencies are very different. Measurement of the bulk and the surface plasmon frequency on the amorphous carbon foil has been extensively carried out recently [22]. We have adopted the formula given in Ref. [18] for the surface plasmon here. This formula, taking the surface electron density from Ref. [22], yields a surface stopping power that varies with ion velocities as shown in Table I. Using the estimated stopping power and the measured SELF energy loss we have evaluated the mean range of the potential as can be seen in Table I also. We note (Table I) that the range varies from 30–60 Å for PP-Al, 40–80 Å for PET-Ge and the similar trend for PP-Au for heavy ions, which is consistent with the surface potential range as predicted by Echenique and Pendry [23]. In contrast the H ions show a large range.

Since the Ge layer was much thicker the theoretical energy loss difference between Ge-PET and PET-Ge is more, and for low energy ions is nearly 200 keV. Again the energy loss in Ge-PET is larger than PET-Ge. In contrast, we observe an energy loss greater in PET-Ge. The measured SELF differential energy loss differences are in the range of 200 to 400 keV in PET-Ge. The Au layer thickness in PP-Au was the smallest 15 nm only and the observed energy loss is found to be the lowest. Hence, we note from Table I that the SELF energy loss depends on the metal layer thickness. The SELF energy loss varies from 100–400 keV, depending on beam energy, on bilayer target materials, and in particular, on the metal layer thickness.

The charge state of incident ions plays a considerable role in preequilibrium energy loss [24,25]; however, the current experiment deals with the equilibrium energy loss in the solid target. Higher beam energies produce higher charge states; however, the measured energy loss differences are higher at the lower beam energies. This observation in fact implies different beam energy dependence.

TABLE III. Differential energy loss straggling in bilayer targets for Fe ions.

| Beam Energy (MeV) | Blank | FWHM (MeV) | | Straggling $\Omega^2$ (MeV$^2$) | | Difference in $\Omega^2$ (MeV$^2$) |
|---|---|---|---|---|---|---|
| | | PP-Al | Al-PP | PP-Al | Al-PP | |
| $86.908 \pm 0.005$ | $1.306 \pm 0.052$ | $2.370 \pm 0.038$ | $2.271 \pm 0.032$ | $3.912 \pm 0.029$ | $3.451 \pm 0.022$ | $0.461 \pm 0.037$ |
| $154.473 \pm 0.006$ | $1.032 \pm 0.013$ | $1.938 \pm 0.020$ | $1.864 \pm 0.017$ | $2.690 \pm 0.023$ | $2.411 \pm 0.020$ | $0.279 \pm 0.031$ |
| | | PP-Au | Au-PP | PP-Au | Au-PP | |
| $86.908 \pm 0.005$ | $1.384 \pm 0.044$ | $2.449 \pm 0.040$ | $2.393 \pm 0.052$ | $4.079 \pm 0.033$ | $3.812 \pm 0.049$ | $0.267 \pm 0.059$ |
| | | PET-Ge | Ge-PET | PET-Ge | Ge-PET | |
| $86.908 \pm 0.005$ | $1.306 \pm 0.052$ | $2.155 \pm 0.045$ | $2.049 \pm 0.044$ | $2.939 \pm 0.043$ | $2.491 \pm 0.045$ | $0.448 \pm 0.062$ |

Finally, we made an attempt to substantiate the effect of SELF through the energy loss straggling [26]. We see here that the SELF introduces 7%–18% additional straggling when the beam is exiting from metal layers as shown in Table III, the thicker metal layer exhibits a larger difference. In this case again theoretically the bilayer target with an insulator on the exit side predicts larger straggling; however, we observe the opposite; details will be reported elsewhere.

To summarize, our measurements of the stopping powers of fast, highly charged iron ions passing through thin bilayer targets indicate an initially surprising differential energy loss when the bilayers are reversed. We find the energy loss depends on the ordering of the target (PP-Al or Al-PP) and is significantly different for the two cases. This energy loss is greater when the metal part is the exit foil section. The energy loss straggling data corroborate the differential energy loss data. Since the exit wake-field for the metal part is predicted to be larger than that for polypropylene or PET (a nonconductor), we propose that the differential energy loss as well as the differential straggling is due to this exit wake field. These measurements reveal wake-field-induced Stark mixing of the substates in a solid [9,10]. Hence, we tentatively ascribe this extra differential energy loss to the stronger metal exit wake field.

Our differential energy loss results show reasonably close agreement with theoretical estimates, from the work of Vager and Gemmell [18]. Interestingly, the mean range of the wake potential is in accord with the prediction by Echenique and Pendry [23]. As metal insulator and metal semiconductor bilayers find potential applications [16], further measurements with bilayer foils of differing thicknesses and conductivity, and especially of the Coulomb explosions of molecular projectiles, can be expected to help develop a consistent theoretical understanding of the origin of the differential energy loss, and verify our proposed SELF wake field hypothesis.

The authors acknowledge the Pelletron crew for providing high-quality beams and Dr. Amit Roy, director IUAC, for providing a lot of beam time from his discretionary quota. We also thank Helmut Weick, R. W. Dunford, A. P. Mishra, Samit Mandal, F. Singh, and Y. Azuma for valuable inputs, and Bhawna Arora, Anil Batra, Gaurav Sharma, and Basu Kumar for their help during the experiments. Very special thanks to A. Kothari and P. Barua for the beam line adjustments, to S. R. Abhilash for target fabrication; also, special thanks to U. G. Naik for providing the clean electric grounds and V. V. V. Sathyanarayana, P. Sugathan, and E. T. Subramaniam for the data acquisition.

*************